\documentclass[graybox, envcountchap]{svmult}

\usepackage{mathptmx}        % selects Times Roman as basic font
\usepackage{helvet}          % selects Helvetica as sans-serif font
\usepackage{courier}         % selects Courier as typewriter font
\usepackage{makeidx}         % allows index generation
\usepackage{graphicx}        % standard LaTeX graphics tool
\usepackage{multicol}        % used for the two-column index
\usepackage[bottom]{footmisc}% places footnotes at page bottom

\makeindex             % used for the subject index
\usepackage{amsmath}
\usepackage{amssymb}
\usepackage{lscape}
\usepackage{multirow}

\def\gapp{\ifmmode\stackrel{>}{_{\sim}}\else$\stackrel{<}{_{\sim}}$\fi}
\def\gsim{\lower.5ex\hbox{\gtsima}}
\def\gtsima{$\; \buildrel > \over \sim \;$}
\def\lapp{\ifmmode\stackrel{<}{_{\sim}}\else$\stackrel{<}{_{\sim}}$\fi}
\def\lsim{\lower.5ex\hbox{\ltsima}}
\def\ltsima{$\; \buildrel < \over \sim \;$}

\newcommand\apgt{\ {\raise-.5ex\hbox{$\buildrel>\over\sim$}}\ }
\newcommand\aplt{\ {\raise-.5ex\hbox{$\buildrel<\over\sim$}}\ }
\begin{document}
\pagestyle{empty}
\frontmatter%%%%%%%%%%%%%%%%%%%%%%%%%%%%%%%%%%%%%%%%%%%%%%%%%%%%%%

\mainmatter%%%%%%%%%%%%%%%%%%%%%%%%%%%%%%%%%%%%%%%%%%%%%%%%%%%%%%%
%\include{part}
%\include{author}

%%%%%%%%%%%%%%%%%%%% author.tex %%%%%%%%%%%%%%%%%%%%%%%%%%%%%%%%%%%
%
% sample root file for your "contribution" to a contributed volume
%
% Use this file as a template for your own input.
%
%%%%%%%%%%%%%%%% Springer %%%%%%%%%%%%%%%%%%%%%%%%%%%%%%%%%%

%%%%%%%%%%%%%%%%%%%%%%%%%%%%%%%%%%%%%%%%%%%%%%%%%%%%%%%%%%%%%%%%%%%%%%%%%%%%%%%%%%%%%%%%%

%\begin{document}
\setcounter{chapter}{0}

\title{Introduction to the Theory of Stellar Evolution}
% Use \titlerunning{Short Title} for an abbreviated version of
% your contribution title if the original one is too long
\author{Giacomo Beccari and Giovanni Carraro}
% Use \authorrunning{Short Title} for an abbreviated version of
% your contribution title if the original one is too long
\institute{Giacomo Beccari 
and Giovanni Carraro \at European Southern Observatory, Alonso de Cordova 3107, Santiago de Chile, Chile,\\ \email{gbeccari@eso.org, gcarraro@eso.org}}
%\and Name of Second Author \at Name, Address of Institute \email{name@email.address}}
%
% Use the package "url.sty" to avoid
% problems with special characters
% used in your e-mail or web address
%
\maketitle
\label{Chapter:Beccari}

\abstract*{Stars form from the collapse and fragmentation of molecular clouds. After this first stage
of formation and evolution as a pre-main-sequence object, a star begins its life in
the main sequence stage through the ignition of Hydrogen in its core.
From here after, the life of a star is simply described as the evolution of a gaseous mass
through well defined stages of equilibrium between gravitational energy and the energy produced 
by the nuclear reactions in its interior. Depending on its initial
mass, heavier and heavier chemical elements are processed in the stellar core up to iron for the most massive stars.
Finally, the star will end its life by simply cooling down as a white dwarf or exploding as a supernova, 
depending once again on its initial mass.
In this introductory chapter, we briefly review the most important aspects concerning the evolution and nucleosynthesis of 
single stars, form the early stage of its formation to the final stages and death. }

%\abstract{TBD}

\section{The Pre-Main Sequence Phase}

\begin{figure}[!htb]
 \centering%
 \includegraphics[width=119mm]{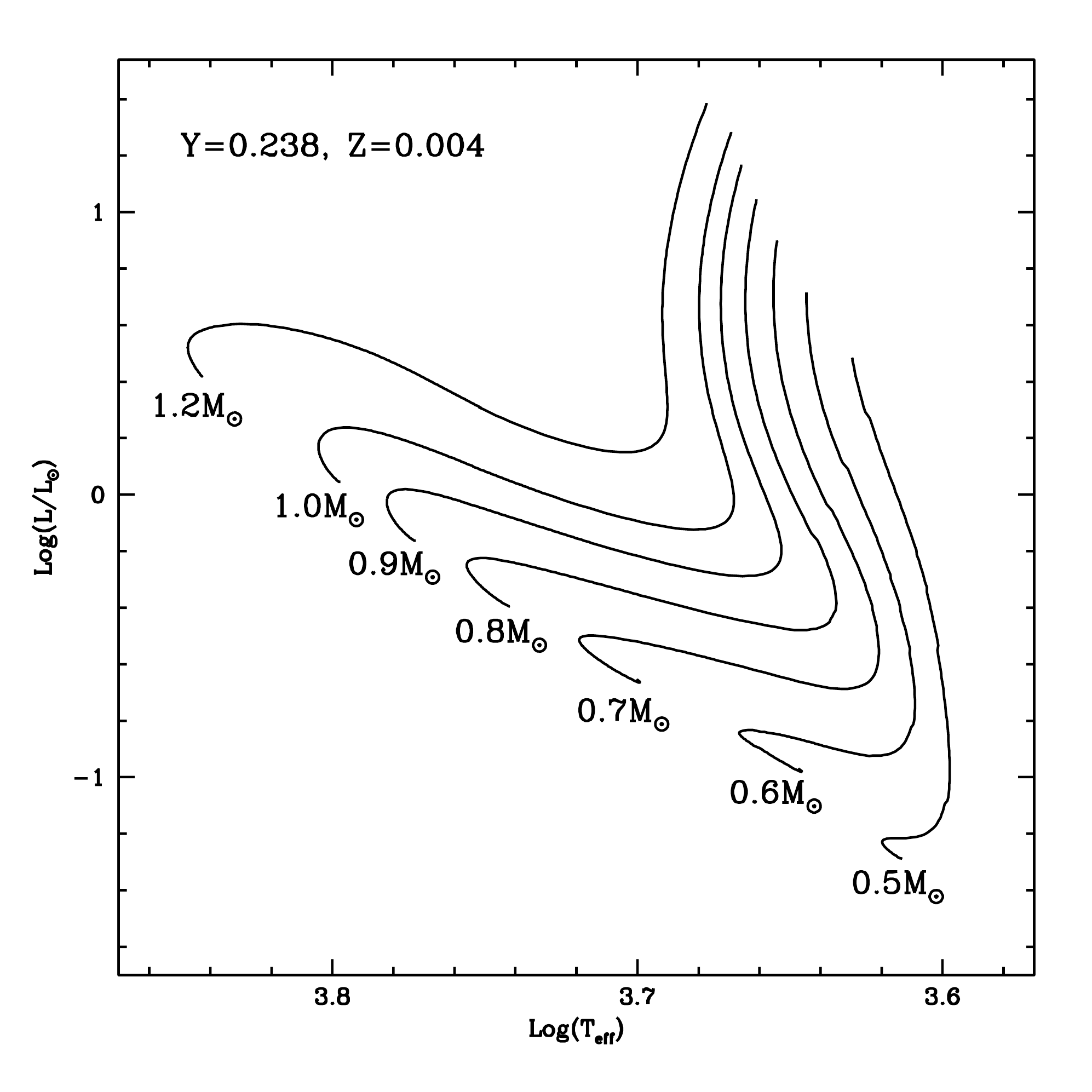}
 \caption{\small A set of Pre-Main Sequence tracks in a HR diagram (theoretical models from \cite{to11}) . The masses of the PMS objects are reported in the figure.}\label{Fig:Htracks}
\end{figure}

Stars form from the collapse and the fragmentation of molecular clouds. This is possible under physical conditions that 
induce a local instability in the cloud. In term of mass, the limit of stability is called Jeans mass.
This critical mass depends on the thermodynamic variables described by the equation:\begin{equation}
 M_{Jeans} \propto T^{3/2}\rho^{-1/2} ,
\end{equation}
making the cold and dense clouds of molecular hydrogen the ideal environment in which the gravitational collapse occurs. \\
In reality, the collapse of a cloud with a  mass of few hundred solar masses is responsible of the formation of stellar associations.  
The following discussion focuses on the behaviour of a single star.\\
The outcome of the collapse of the molecular gas is a fully convective protostar in hydrostatic equilibrium. 
As such, it compensates the lose of energy due to radiation by contracting, following the virial theorem. 
The contraction causes a temperature rise in the core of the structure.
As a consequence the opacity $\kappa$ (the laws of Kramers describe a dependency of type $\kappa$ $\propto$ $T^{-3.5}$) and the 
module of the radiative gradient decrease. The latter is defined as:
\begin{equation}
 \frac{dT}{dr} =- \frac{3}{4ac} \frac{\kappa \rho F}{T^{3}} ,
\end{equation}
where %$\frac{3}{4ac}$ is defined as a constant, 
$F$ is the flux emitted and $\rho$ is the density of the structure.\\
The Schwarzschild criterion states that the convection is triggered when the radiative gradient is greater in magnitude, 
than the adiabatic lapse rate. Hence, the effect of the irradiation of the protostar is to gradually erase the convection inside its structure.\\
In the Hertzsprung-Russell (HR) diagram\index{Hertzsprung-Russell diagram}, i.e. the log $T_{eff}$ -- log $L$ diagram, where $T_{eff}$ is the surface temperature of a star and $L$ its bolometric luminosity,
this results in the protostar departure from the so-called \textit {Hayashi track}. The latter is defined as the evolutionary path traced by
a fully convective structure in hydrostatic equilibrium.\\
At this point, the protostar continues to contract and the temperature in the nucleus continues to increase until it becomes 
sufficiently high ($T\simeq10^{7}$ K) to trigger
the reactions of nuclear burning of hydrogen. At this stage the star reaches the Zero Age Main Sequence\index{Zero Age Main Sequence} (ZAMS, see Fig.\ref{Fig:Htracks}).\\
It is important to know that $M_{lim}\simeq0.08$ M$_\odot$ represents the mass limit below which the 
physical conditions needed for the activation of nuclear reactions are not reached. Hence, this value represents the minimum 
mass for an object to become a star. Any object of mass lower than $M_{lim}$ will be a brown dwarf or a planet.

\section{The Main Sequence}

\begin{figure}[!htb]
 \centering
 \includegraphics[width=119mm]{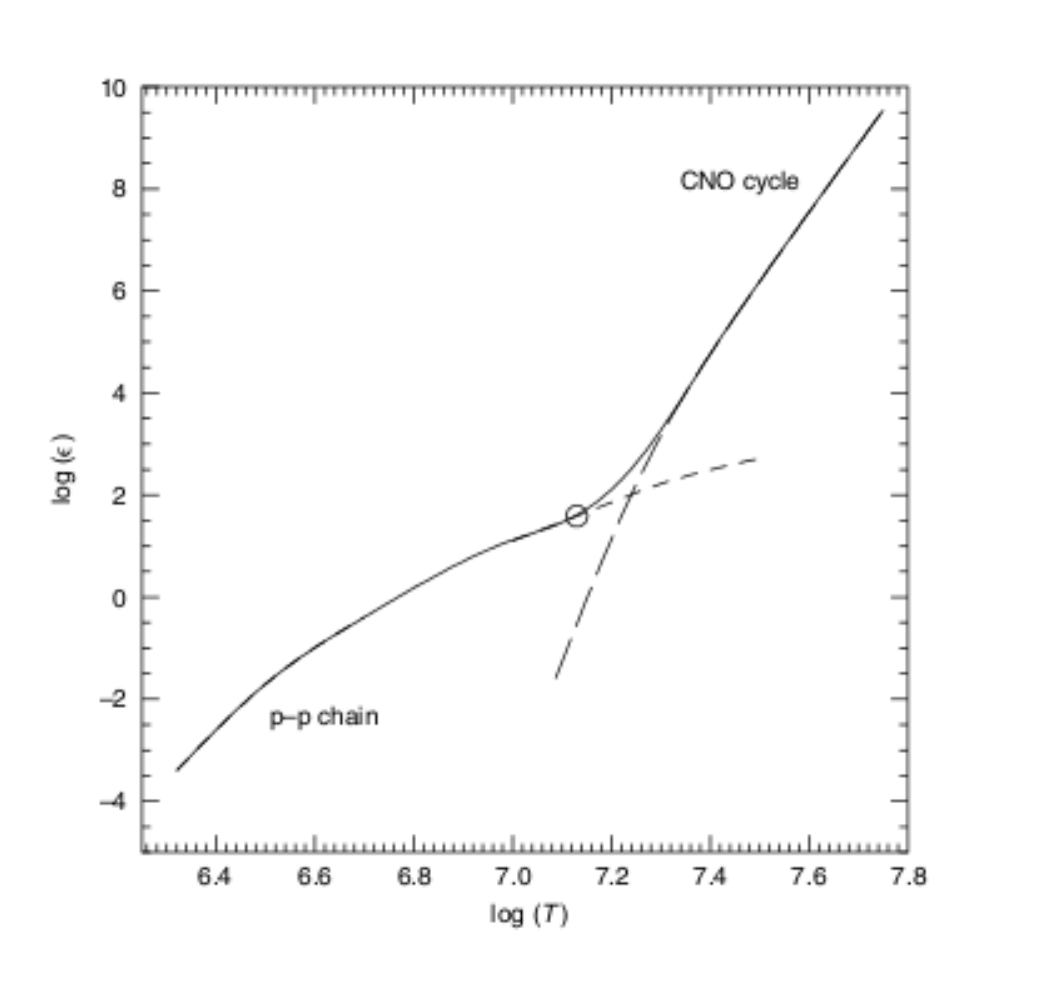}
 \caption{\small Efficiency of the PP chain and CNO cycle as a function of the temperature. The picture is reproduced from \cite{sa05} with
  permission from the authors.}\label{Fig:cno}
\end{figure}

The Main Sequence\index{main sequence} (MS) is  defined as the stage during which a star burns hydrogen in its nucleus.
This is the longest evolutionary phase in the life of a star. 
Such burning may occur through two different mechanisms: either through the proton-proton (PP) chain\index{p-p chain}, or through the 
Carbon-Nitrogen-Oxigen (CNO) cycle\index{CNO cycle}. The two types of combustion have very different impacts on the stellar structure: 
stars that produce energy via the PP chain, in fact, are characterised by a radiative core and a convective envelope, 
while stars that consume hydrogen through the CNO cycle have a convective core and a radiative envelope.\\
Both processes consume four hydrogen nuclei to give as net product an helium nucleus. As a consequence 
the energetic budget involved in both processes is very similar.
Nevertheless, from the equations:
\begin{equation}
\label{eq_pp}
 \varepsilon_{pp} \propto \rho X^{2} T_{6}^{4} 
\end{equation}
\begin{equation}
\label{eq_cno}
 \varepsilon_{CNO} \propto \rho X T_{6}^{16} ,
\end{equation}
where $X$ is the mass fraction of hydrogen and $T_{6}$ is the temperature expressed in millions of Kelvin, it is
clear that the reactions have very different temperature dependences.  Hence, at lower temperatures the 
PP chain dominates, but with rising temperatures there is a sudden transition to dominance by the CNO cycle, which 
has an energy production rate that varies strongly with temperature. This is why the CNO cycle is more important 
for more massive stars: their interior temperatures are higher, thus favouring the CNO cycle.\\
Moreover, since:
\begin{equation}
 \frac{dL}{dr} = 4 \pi r^{2} \rho (r) \varepsilon ,
\end{equation}
during the CNO cycle the luminosity and hence the flux emitted is very high and, as a consequence of the
Schwarzschild criterion, the convention is activated in the stellar nucleus.
The opposite happens in the case of a PP chain.\\
Typically, the value of mass that separates these two behaviours is $\sim1.2 $M$_{\odot}$. It should however be noted that, 
in most cases both burning channels are active. The distinction in the two cases simply describes the predominance of one channel over the other, depending on the temperature that is reached in the core (see Fig.~\ref{Fig:cno}).\\

The entire evolution of a star on the MS is then characterised by the consumption of hydrogen in the core
with the production of helium. As a consequence, the average molecular weight, defined as
\begin{equation}
 \mu=\frac{1}{(2X+0.75Y+0.5Z)} ,
\end{equation}
where $Y$ is the helium mass fraction and $Z$ the mass fraction of all the other elements\footnote{$Z$ is called in astronomy the \emph{metallicity}, and is given by $Z=1-X-Y$.}increases. As a result, considering as valid the approximation of non-degenerate gas (necessary to build stable cycles of thermonuclear reactions), 
the pressure decreases and the core slowly contracts. Both the density and the temperature in the core --- and thus also the 
efficiency of the nuclear reactions --- increase as a consequence of the contraction. The effect of all this is, again, 
an increase of the flux emitted, with the consequent introduction of convective motions in the stellar interiors. 
It is in this way, then, that the star starts its ascent on the subgiant\index{subgiant} branch (SGB).

\section{The Combustion of Hydrogen in a Shell: the Sub and Red Giant Branches}

\begin{figure}[!htb]
 \centering%
 \includegraphics[width=119mm]{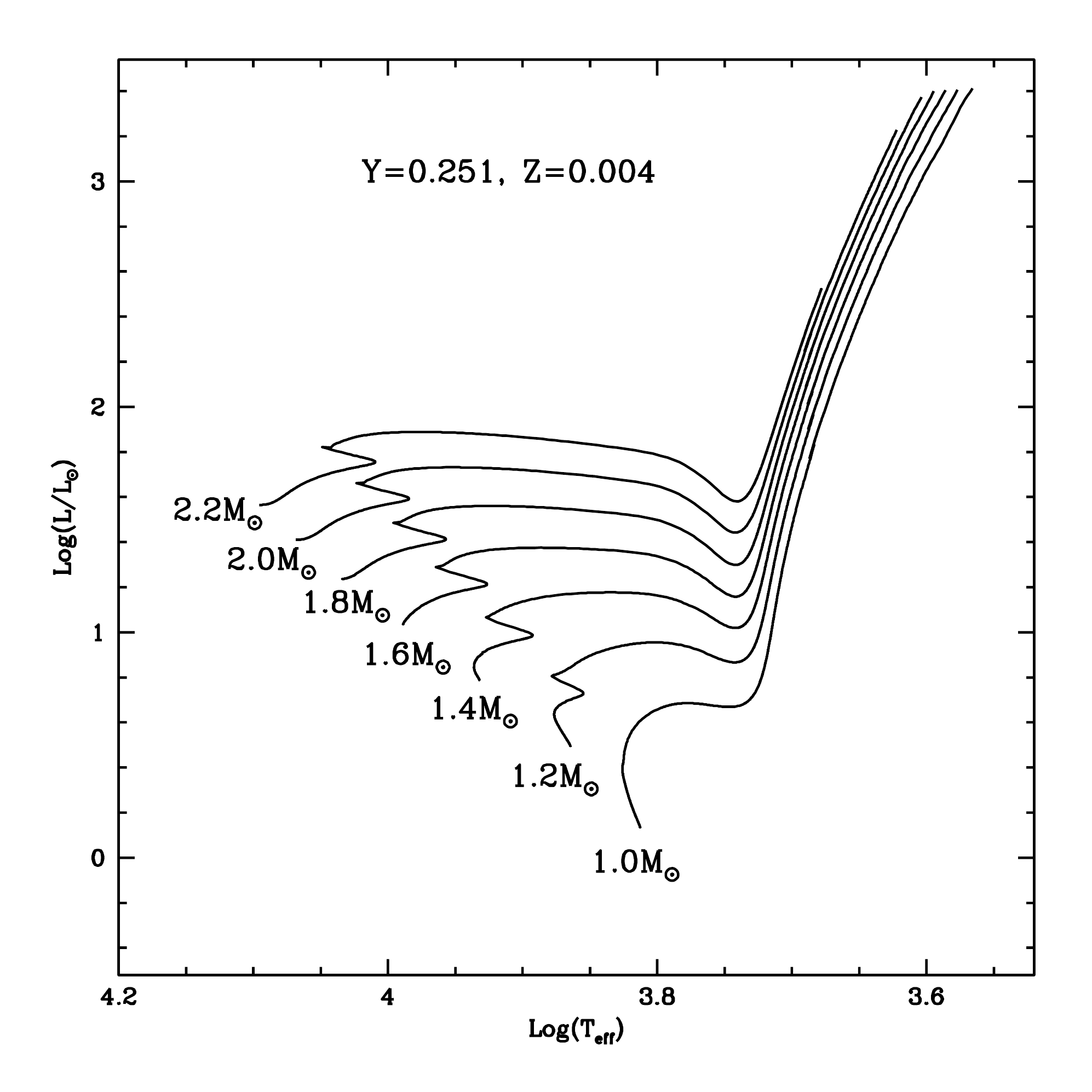}
 \caption{\small Evolutionary tracks at a given metallicity showing the SGB and RGB phases for
 stars at different masses (as indicated in the Figure; theoretical models from \cite{pi06}).}\label{Fig:tracks}
\end{figure}

The evolutionary path of a star after the consumption of the hydrogen in the core depends of its mass.\\
The first important distinction to be considered is that between stars of mass less than or greater than  $\sim 1.2$ M$_\odot$ or, 
as previously described, between stars with or without a convective core.\\
The effect of convection is to smooth any gradients of chemical abundances in the region where it is active. 
This means that the hydrogen in the radiative nuclei is exhausted only in the innermost region, while in the regions 
immediately adjacent there are already the conditions to trigger the reactions of hydrogen burning in thick shells.  
Convective nuclei are characterised by the presence of extensive regions, above the nucleus itself, in which the availability of fuel is low. Before it can ignite the hydrogen burning in the shell, which in this case is more distant and colder, 
the structure must contract and heat up to a temperature of $\sim10^7$K.\\
This translates along the evolutionary tracks on the HR diagram, in a continuous sequence in the case of low mass stars and in a broken line, characterised by a ``hook'', for stars in which the convection is dominant (Fig.~\ref{Fig:tracks}). 
During this phase the gravitational energy is the only source of energy .\\
Once turned on, the shell becomes the main energy source of the star in terms of luminosity:  a nucleus of inert helium  has formed
at the centre as the temperatures are not high enough to activate the helium burning (T $\sim 1.5\cdot10^{8}$ K).\\ 
However, the shell proceeds consuming hydrogen outward of the stellar structure, and as consequence, inert helium 
is deposited in the core. If the core is in conditions of perfect gas, it remains isothermal until its mass does not exceed 
the critical value of the mass of Schonberg-Chandrasekhar (M$_{SC}$): when this occurs, or when $M_{core}>0.12 M_{star}$, the core shrinks until the activation of the helium-burning reaction, also called reaction 3$\alpha$.\\
As the shell keeps on burning hydrogen along the stellar structure, it also gets progressively thinner. The structure responds by
expanding and cooling at the surface. At this stage the stars move in the HR diagram to the regions dominated by convection, 
which is indeed penetrating the stellar volume from the outside. When the star is sufficiently close to the Hayashi track, the SGB phase 
ends.\\

The stellar structure is not able to reach and/or overcome the Hayashi track which is by definition suitable for fully convective structures.
The structure responds to the solicitations from the shell by rapidly increasing its luminosity and its radius (Fig.~\ref{Fig:tracks}). 
As a consequence the temperature at the surface decreases, and thus it begins the Red Giant Branch\index{red giant} (RGB) phase.

\begin{figure}[!thb]
 \centering%
 \includegraphics[width=119mm]{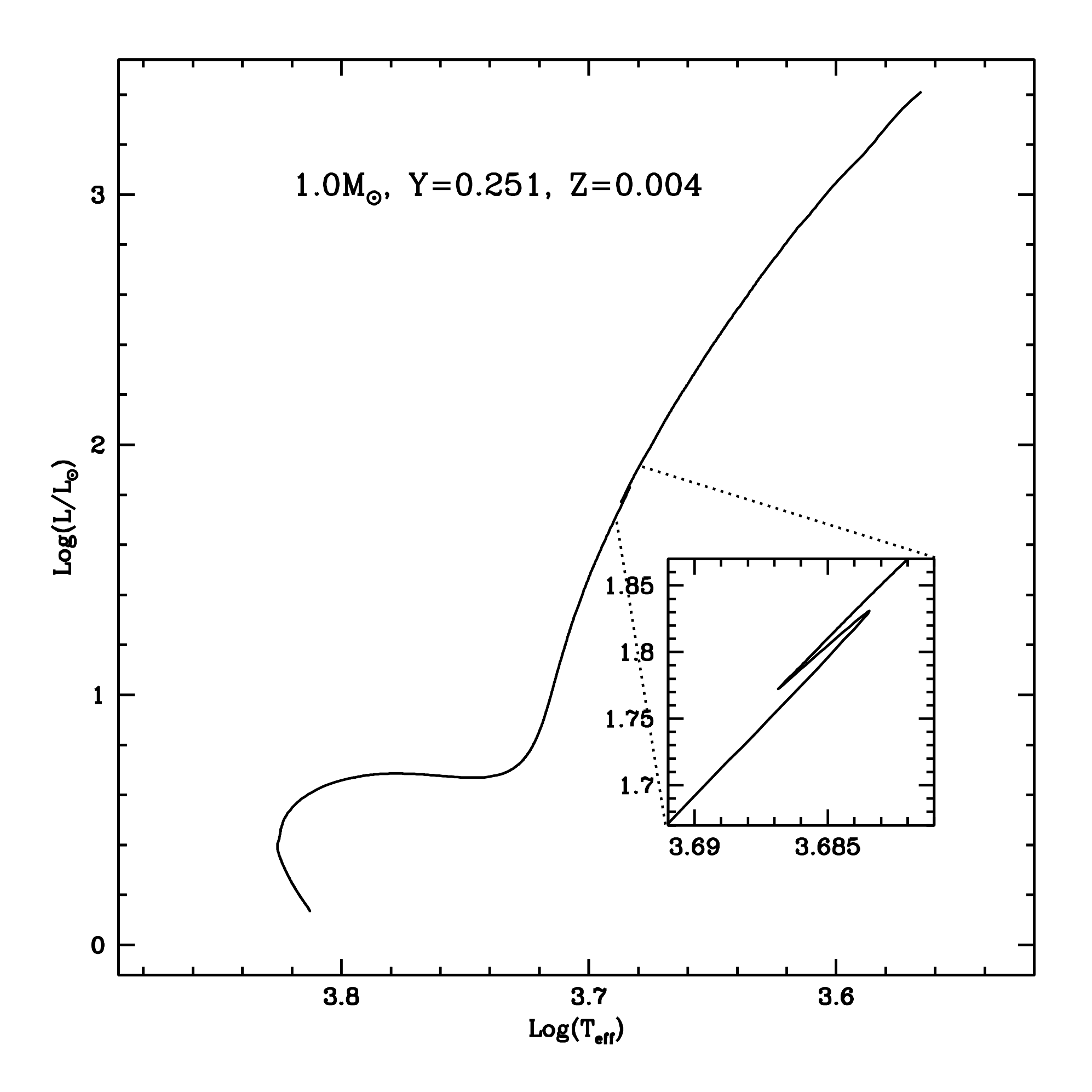}
 \caption{\small The behaviour of an evolutionary track at the RGB bump is show in the inset box together with a 1M$_{\odot}$ track (theoretical model from \cite{pi06}).}\label{Fig:bump}
\end{figure}

It is important to notice that the time needed for a star to reach this evolutionary stage depends on the initial mass of the star itself.
In particular, if a star is very massive ($M>6$ M$_\odot$), the mass of its core at the end of the MS is already greater 
than $M_{SC}$,  and thus the duration of the phase of RGB is extremely short.\\
For intermediate massive stars (2.2 M$_\odot<M<6$ M$_\odot$), the RGB phase is slower because it is necessary 
that the shell generates enough helium to exceed the value of critical mass  and the core can start to contract.\\
The situation is different if the star has initial mass $<2.2$ M$_\odot$: in this case, in fact, the structure develops a core only 
partially degenerate that continues to increase its mass while remaining practically isothermal (shrinks very slowly).
This, of course, slows down the nuclear heating and indeed the nucleus is able to trigger the 3$\alpha$ reaction 
only after an extremely long time.\\
Shortly after the beginning of the RGB,  the first event of chemical mixing experienced by a star after the initiation of nuclear reactions
takes place: the so called first \textit{dredge-up}. The convection,  penetrating into the areas affected by the CNO cycle, brings some of its products
to the surface, causing the alterations of the chemical composition. In particular, the surface abundances of He and N increase, while those of C and O decrease proportionally to the characteristic times of nuclear reactions in the CNO cycle.\\
From this point onwards the shell starts to approach the limit reached by the convection which tends to recede. 
A discontinuity in the profile of hydrogen remains in correspondence with this limit, which results in an important characteristic along 
the evolutionary track: the so called \textit{RGB bump}\index{RGB bump}. When the shell, in fact, is located in proximity of the discontinuity, it encounters a region in which the average molecular weight $\mu$ is suddenly lower; since the luminosity of the shell, $L_{shell}$ obeys the relation
\begin{equation}
 L_{shell} \propto T \propto \mu^{7.5} ,
\end{equation}
then a reverse is visible in the evolutionary track on the HR diagram. The RGB stars then move back along the evolutionary track once the shell leaves the discontinuity
behind (Fig.~\ref{Fig:bump}). 

This theoretical sequence is observationally witnessed by an increase of density of stars in a small region along the RGB, created 
by the fact that a star spends a relatively long time in a short range of luminosity and temperature. \\
The luminosity of the RGB bump is therefore a critical observable allowing one to understand how deep the convection pushed into the stellar interiors.\\

After the bump, the stars move very rapidly along the RGB up to the last point of this evolutionary sequence: the \textit{RGB tip}\index{RGB tip}.\\
It is important to emphasise again the golden rule that the initial mass of the star is driving the evolution of the star itself along these firsts evolutionary stages: 
massive stars ($M>15$ M$_\odot$), for example, do not experience the phase of RGB, as the core is already sufficiently massive to contract  
and warm up to the quick triggering of the 3$\alpha$ reaction. The stars of intermediate mass (2 M$_\odot<M<15$ M$_\odot$) develop 
a short RGB, so that it is not theoretically expected for the stars to pass to the stage of RGB bump. The core, in fact, reaches a temperature of 
$T\simeq1.5\cdot10^{8}$ K, required to trigger the helium-burning before the shell reaches the discontinuity of the profile of hydrogen. 
For low mass stars ($M<2.2$ M$_\odot$), finally, the RGB has the greatest importance. In fact, it is widespread in brightness and 
lasts longer, which results in a very populated and bright giant branch in the observed colour-magnitude diagram\index{colour-magnitude diagram}.

\section{The Helium-Burning in the Core: the Horizontal Branch}

\begin{figure}[!thb]
 \centering%
 \includegraphics[width=119mm]{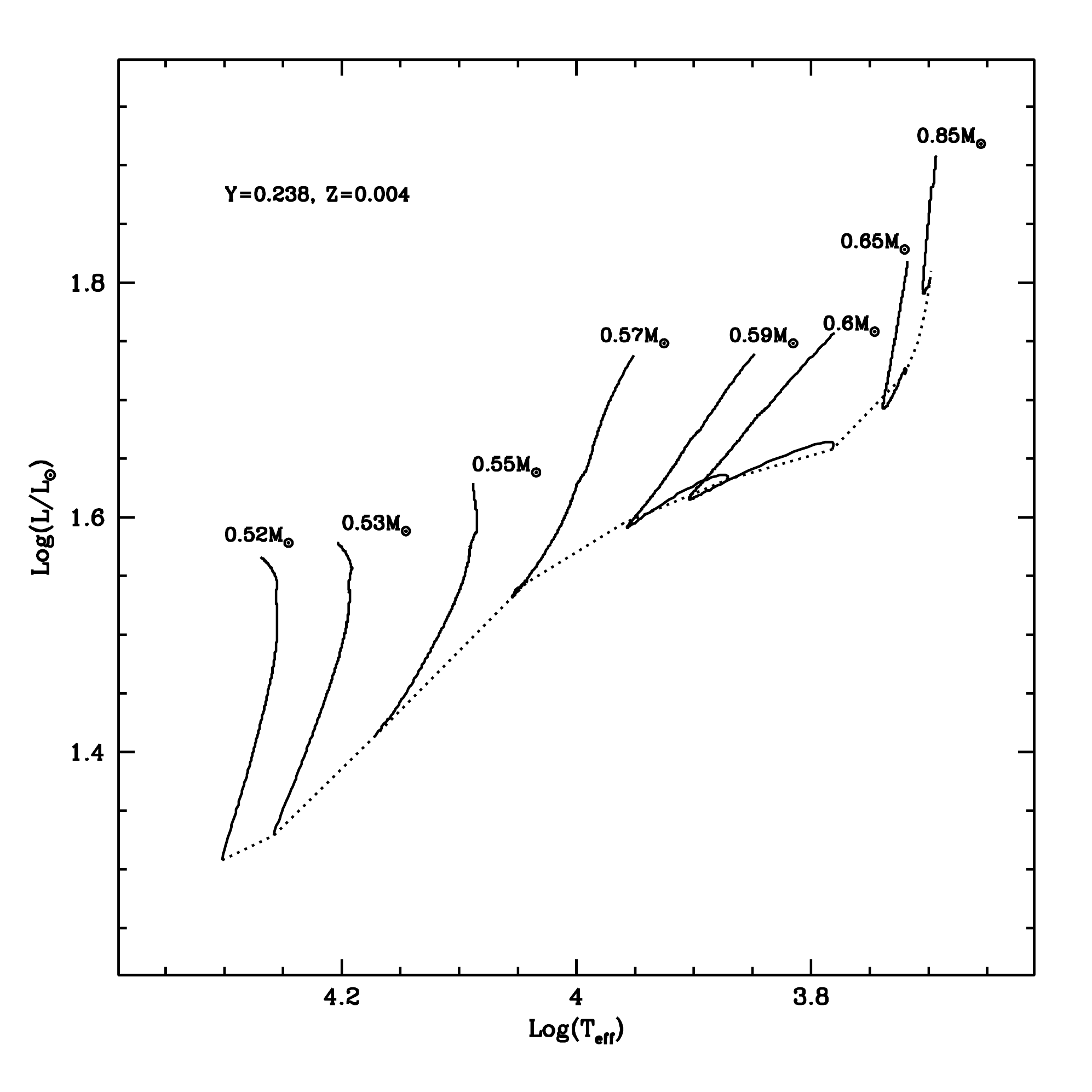}
 \caption{\small Theoretical ZAHB (dashed line) with a progenitor of $M=0.9$M$_{\odot}$. The models are taken from \cite{ca03}.}\label{Fig:zahbs}
\end{figure}

Keeping in mind the distinction on the initial mass proposed in the previous section, it is clear that
for massive stars the fusion of the helium core happens rapidly and in a continuous manner with respect to the previous stage of burning, without 
producing important features on evolutionary tracks. The situation is very different for the intermediate and low-mass stars.\\
As described above, because of the combustion of hydrogen in the core, stars of initial mass $<2.2 $M$_\odot$ 
develop a semi-degenerate core, which accretes  mass from the shell during the RGB phase without contracting and remaining isothermal. 
The partial degeneration of the environment has two important effects: the temperature of the combustion of helium is lowered and the ability 
of contraction of the core decreases, delaying the triggering of  the $3\alpha$ reaction. Moreover, the initiation of the $3\alpha$ reaction in a
degenerated environment causes a thermonuclear runaway. This is the consequence of the lack of the negative feedback between
temperature and pressure, as (in non-relativistic degenerate environment):
\begin{equation}
 P \propto \rho^{5/3}. 
\end{equation}
The ignition of the helium burning is, for these reasons, semi-explosive and is generally referred to as the \textit{He-flash}\index{helium flash}.\\
This evolutionary stage is defined as a flash since the release of energy is almost instantaneous and sizable, 
corresponding to about $10^{11}$ L$_\odot$. Nevertheless, this energy is used for the expansion of the semi-degenerate stellar structure
and for the removal of the degeneration itself.\\
In particular, the helium flash  is nothing but a series of flash that are repeated until the degeneration of the nucleus is
completely removed and the stable burning of helium is performed in conditions of perfect gas.\\
In detail, the flash will originate in the following way: first, the star that is evolving along the RGB is characterised by 
conditions of density and temperature such as to enable the interaction between the photons produced in 
the nucleus and the degenerate electrons, according to the relation
\begin{equation}
 \gamma + e^{-} \longrightarrow e^{-} + \nu + \bar{\nu} .
\end{equation}
The neutrinos produced, having a small cross section, tend to leave the structure subtracting energy. 
In particular the inner regions tend to cool more rapidly and, consequently, in the core of the star forms a reverse 
gradient of temperature that causes the ignition of the first flash in the outer area of the core. 
The sudden release of energy massively increases the flux emitted and the convective motions that are established 
expand the structure by partly removing the degeneracy. 
What remains is a small semi-degenerate core that undergoes the same process repeatedly until the degeneracy is 
removed completely and creates the conditions that allow the stable process of helium burning in a perfect gas.\\
It is important to specify that the He-flash is still not the phase of helium burning: only 5\% of that element is consumed during the flash.\\
The stable helium burning in the core according to the reaction $3\alpha$
\begin{equation}
{\rm ^{4}He+^{4}He \longrightarrow ^{8}Be  }
\end{equation}
\begin{equation}
{\rm  ^{8}Be+^{4}He \longrightarrow ^{12}C +\gamma,}
\end{equation}
starts along the so called \textit{Zero Age Horizontal Branch} (ZAHB)\index{horizontal branch} defined, in the HR diagram, as the locus of points in which a star 
is positioned, in a very short time ($\sim10^6$ yr), 
once the helium flash is finished. Notice that the ZAHB is located on the HR
diagram at a lower luminosity with respect to the tip of the RGB, as a consequence of the
expansion and relaxation of the nucleus of the stars after the He-flash.\\
It should be noted that for low mass stars the helium flash happens when the mass of the degenerate core is 0.5 M$_\odot$. 
This is a key ingredient to understand where a star positions itself on the ZAHB.\\
For stars of intermediate mass, however, the core is not in degenerate conditions and turns on the reactions of helium-burning 
once it exceeds the mass M$_{SC}$ and when the ignition temperatures are reached.  Even in this case, however, the evolutionary 
tracks decrease in luminosity with respect to the end point of the RGB. The shell, which keeps the dominant role in terms of emission, 
is pushed towards the outer and colder regions, where the efficiency $\epsilon$ decreases.\\  
The positioning of a star on the ZAHB 
depends on the ratio between the mass of the nucleus and the total mass of the star. In this sense, the ZAHB (as for the ZAMS) represents a 
sequence in mass (see Fig.~\ref{Fig:zahbs}).  

Since the low-mass stars reach the tip of the RGB\index{red giant} at constant mass of 
the nucleus, the location of these stars on the HB depends only on the mass of the envelope.  In particular, the higher this value, 
and therefore the greater is the total mass of the star, the redder the point of the ZAHB in which it is positioned. In fact a thicker 
envelope is able to ``shield'' better the star, which then appears cooler and redder. For these same reasons, stars of intermediate 
mass do not describe a real sequence on  the HR diagram: they locate themselves in the reddest part of the ZAHB and constitute 
the so-called \textit{He-clump}\index{helium clump}.\\ 
When the luminosity of the shell dominates with respect to the emission from the nucleus, the evolutionary track moves towards the bluest regions of the colour magnitude diagram. Conversely, if it is the core that dominates, the stellar structure tends to expand, to cool and  the tracks then moves towards the convective 
zone of the HR diagram, i.e. towards the Hayashi's track (Fig.~\ref{Fig:zahbs}).  At the end of the HB phase, when the helium in the core is exhausted, the star moves towards the \textit{Asymptotic Giant Branch} (AGB)\index{asymptotic giant branch star}.

\section{Two burning shells: the AGB}

\begin{figure}[!htb]
 \centering
\includegraphics[width=119mm]{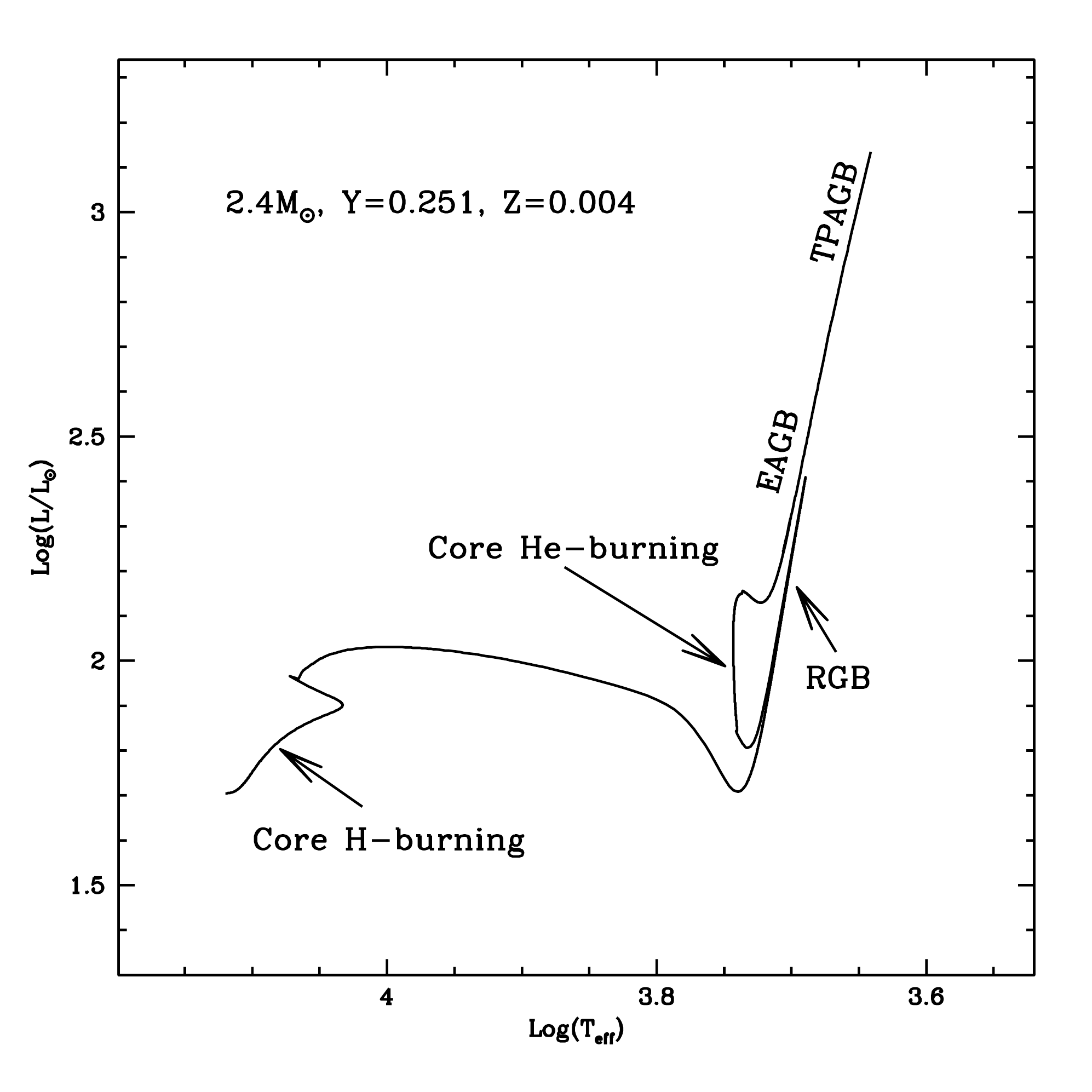}
\caption{\small Theoretical evolutionary track for an intermediate massive star (2.4 M$_{\odot}$; theoretical model from \cite{pi06}).}\label{Fig:agb}
\end{figure}

At the end of the HB phase, the core of the stars, consisting of carbon and oxygen, becomes extremely compact being
formed by a smaller number of atoms, and therefore warmer. The same properties are shared by the adjacent regions, 
and consequently it is possible to turn on the helium-burning reaction in a shell.\\
Once again the trigger occurs in a region partially degenerate and therefore leads to the release of a huge amount 
of energy. The first flash that is generated traces on the HR diagram the so-called stretch of Early AGB 
(EAGB, see Fig.~ \ref{Fig:agb}).\\

During the EAGB the shell of hydrogen, still on, continues to deposit helium in the
layer that separates it from the core of C-O, making it more and more degenerate. As already discussed, the degeneracy 
lowers the ignition temperature of the helium until the shell lights up in a semi-explosive way. The emission of energy, 
this time of approximately $10^{6}$ L$_\odot$, translates in the expansion of the inter-shell that removes the degeneracy 
and pushes the shell of hydrogen towards the outer and colder regions, reducing the luminosity.
The shell of helium becomes dominant, and continues to deposit CO on the core, making it more and more degenerate.
At the same time the inter-shell stops progressively to expand until the shell of hydrogen returns to dominate.\\
During this phase of EAGB, the stars in the mass range $M>4.6$ M$_\odot$, can experience the second chemical mixing process. 
As already mentioned, the expansion of the structure allow the convection layers to penetrate from the outside. 
If this is sufficiently deep to arrive to the inter-shell, it can bring at the surface elements such as He, C, N, O, which are the 
products of combustion of hydrogen. In this sense, the second dredge-up is extremely similar to the first.\\
The next phase is called Thermally Pulsing AGB (TPAGB) and substantially is the cyclic repetition of the phenomena 
that have occurred during the EAGB, which can be just considered as the first of these thermal pulses. 
The fundamental process that occurs during TPAGB is the third dredge-up. Since the convection is penetrating 
from the outside (following the cooling of the envelope) and at the same time it must be expanding from the inside 
(thanks to the thermal pulse), it may happen that, during the inactivity of the shell of hydrogen, the two convective 
streams are in contact. In this case, therefore, the mixing involves extremely profound regions of the star ($\sim75\%$ 
of the structure) and generates chemical signatures that differ form those characterising the other two dredge-ups, 
such as an increase of the abundance of carbon at the surface.\\
During the entire duration of the AGB phase, the stars may be affected by phenomena of substantial mass loss. 
This is due to the fact that the envelope of these stars is expanded and cold as ever during the other phases of evolution: 
therefore the formation of layers of molecules and dust becomes possible, layers which are then removed as stellar wind by the effect of radiation
pressure. %It seems that this process is particularly important for medium-large massive stars, while for the other stars most of the mass loss ($\sim0.2$ M$_\odot$) occurs during the RGB.\\

These phenomena of mass loss become even more important when considered from the point of view of the chemistry of the
medium in which stellar populations evolve. Besides the two dredge-ups, in fact, other peculiar chemical events take place 
during the AGB. One of these is the production of s-process elements\index{s-process element} by neutron capture reactions. During the thermal pulse 
in the intershell, convection remixes isotopes such as $^{14}N$ and $^{4}He$ that may come in contact and react, giving 
rise to nuclear chains such as:
\begin{equation}
{\rm ^{14}N + ^{4}He \longrightarrow ^{18}F + \gamma }
\end{equation}
\begin{equation}
{\rm ^{18}F \longrightarrow ^{18}O + e^{+} + \nu }
\end{equation}
\begin{equation}
{\rm ^{18}O + ^{4}He \longrightarrow ^{22}Ne + \gamma }
\end{equation}
\begin{equation}
{\rm ^{22}Ne + ^{4}He \longrightarrow ^{25}Mg + n.}
\end{equation}

These reactions require temperatures $T>3.5\cdot 10^{8}$ K to complete and only stars of mass $M>3$ M$_\odot$
satisfy such a requirement. Free neutrons that are created are then captured and contribute to form s-elements that are 
quickly brought to the surface thanks to the dredge-up and are then dispersed in the interstellar medium through the 
above-mentioned events of mass loss. They then become the important tracers for AGB stars of this type. 

In lower-mass AGB stars, the intershell region does not reach the required
temperature to start the chemical reaction
involved in this chain. However, in AGB stars with mass $<3$ M$_{\odot}$ another 
reaction that can act as an alternative source of neutrons is $^{13}$C($\alpha$, n)$^{16}$O.
This reaction requires a temperature of $\sim9\cdot10^7$ K to operate.

Another mechanism that can be established during these phases is the so-called \textit{hot bottom burning}: 
stars of mass $M>5$ M$_\odot$ reach, at the base of the convective envelope, temperatures able to trigger the secondary
cycles of the CNO cycle. These proton capture reactions, through the creation of N, O, F, Ne, Na, Mg, may proceed until 
the formation of aluminum, and as for the s-elements, these elements enrich the interstellar medium with proportions
corresponding to the characteristic times of the reaction itself. This event might be at the origin of the peculiar anti-correlations
between some chemical abundances (such as Na and O, or Mg and Al) which characterise all the galactic globular 
clusters\index{globular 
cluster} observed to date.\\

\begin{figure}[!thb]
\centering%
 \includegraphics[width=119mm]{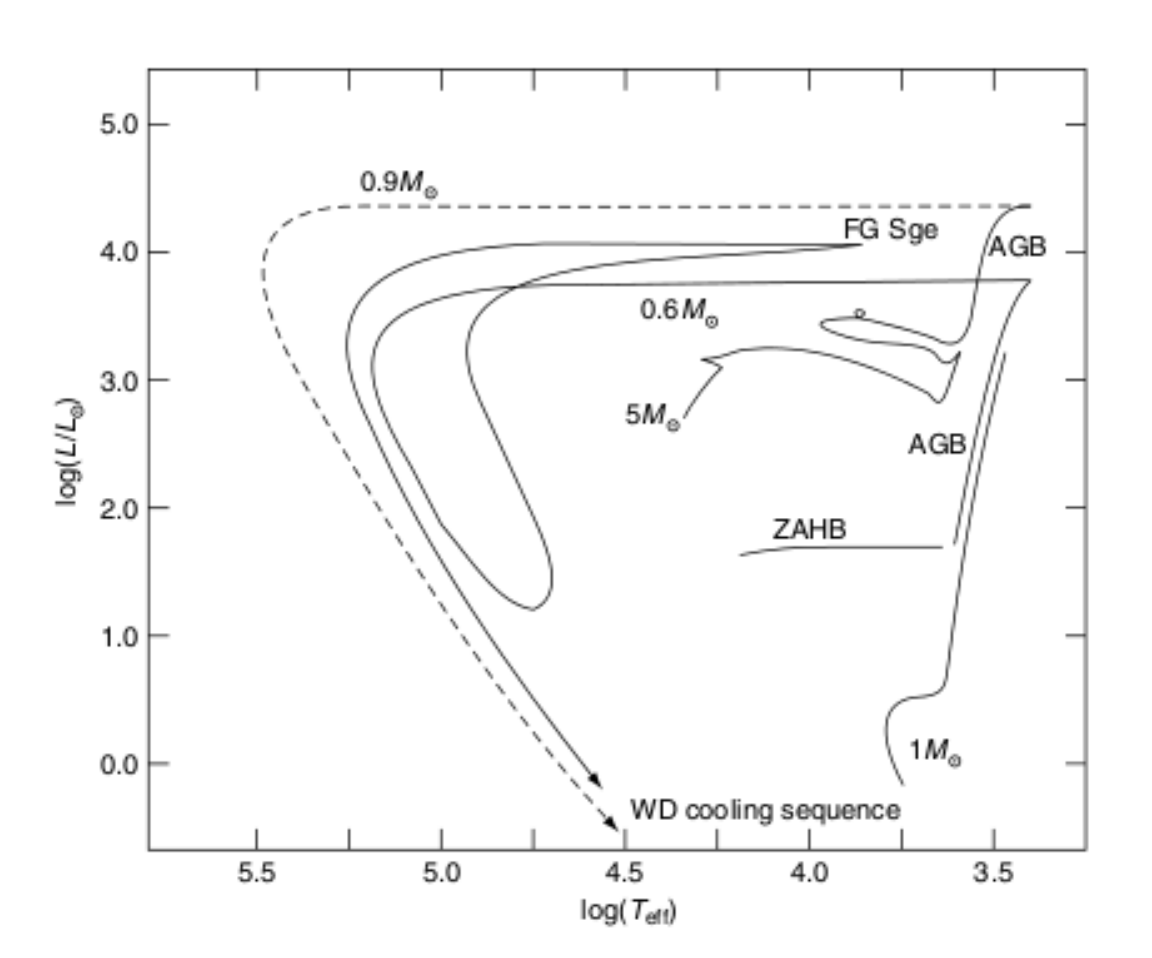}
 \caption{\small Post ZAMS evolutionary tracks for intermediate and low mass stars.  The picture is reproduced from \cite{sa05} with
 the permission of the authors.}\label{Fig:finaktracks}
\end{figure}

After about a dozen thermal pulses, the latest expansion is sufficient to allow the release of the outermost layers around the
nucleus of CO which is rapidly becoming fully degenerate. With this phase, known as post-AGB, the  
nuclear-active evolution of the stars ends leading to the final stages of a star's life (Fig.~\ref{Fig:finaktracks}).

\section{The Final Stages of the Evolution of the Stars}

As for the entire evolution of the star, the latest stages depend on its initial mass.
As for objects of mass less than 0.3 M$_\odot$  (but still greater than the aforementioned limit of 0.08 M$_\odot$),  
once the hydrogen is exhausted and its  burning is finished, they fail to trigger the burning of the helium in the core.
They simply cool as helium white dwarfs. It 's important to specify that objects of this type should not be currently observable,
since such very low mass stars evolve off the MS after about 20 Gyr. However several have been observed and the explanation 
of their nature is linked to phenomena of interaction between two stars in a binary system: when one of the two companions, 
expanding, reaches its Roche lobe\index{Roche lobe} limit, it starts to transfer matter to the secondary. If the primary was evolving 
along the RGB and loses the entire envelope, what remains is just a helium white dwarf\index{helium white dwarf}.\\
Let us now consider stars of mass M between 0.3 M$_\odot<M<8$ M$_\odot$. After the stage of planetary 
nebula\index{planetary 
nebula} --- a brief swan song where the outer shell is ionised by the very hot core, once the outermost layers are dispersed, what remains is a completely degenerate CO. Its destiny is to follow 
the so-called cooling sequence of a white dwarf\index{white dwarf}.
The key feature of a white dwarf is to be in conditions of density and temperature that makes it a structure completely degenerate.
As such, it is supported by the pressure of degenerate electrons. In addition, the mass and radius are linked by a relationship 
of the type:
\begin{equation}
 MR^{1/3}=constant,
\end{equation}
meaning that the more massive a white dwarf is, the smaller it is. Finally, there is a critical value of the mass, called the
\textit{Chandrasekhar mass}\index{Chandrasekhar mass} ($\sim 1.44~{\rm M}_\odot$, see e.g. \cite{sh82}), beyond which the structure is no longer able to maintain equilibrium. 

\begin{figure}[!htb]
 \centering
 \includegraphics[width=119mm]{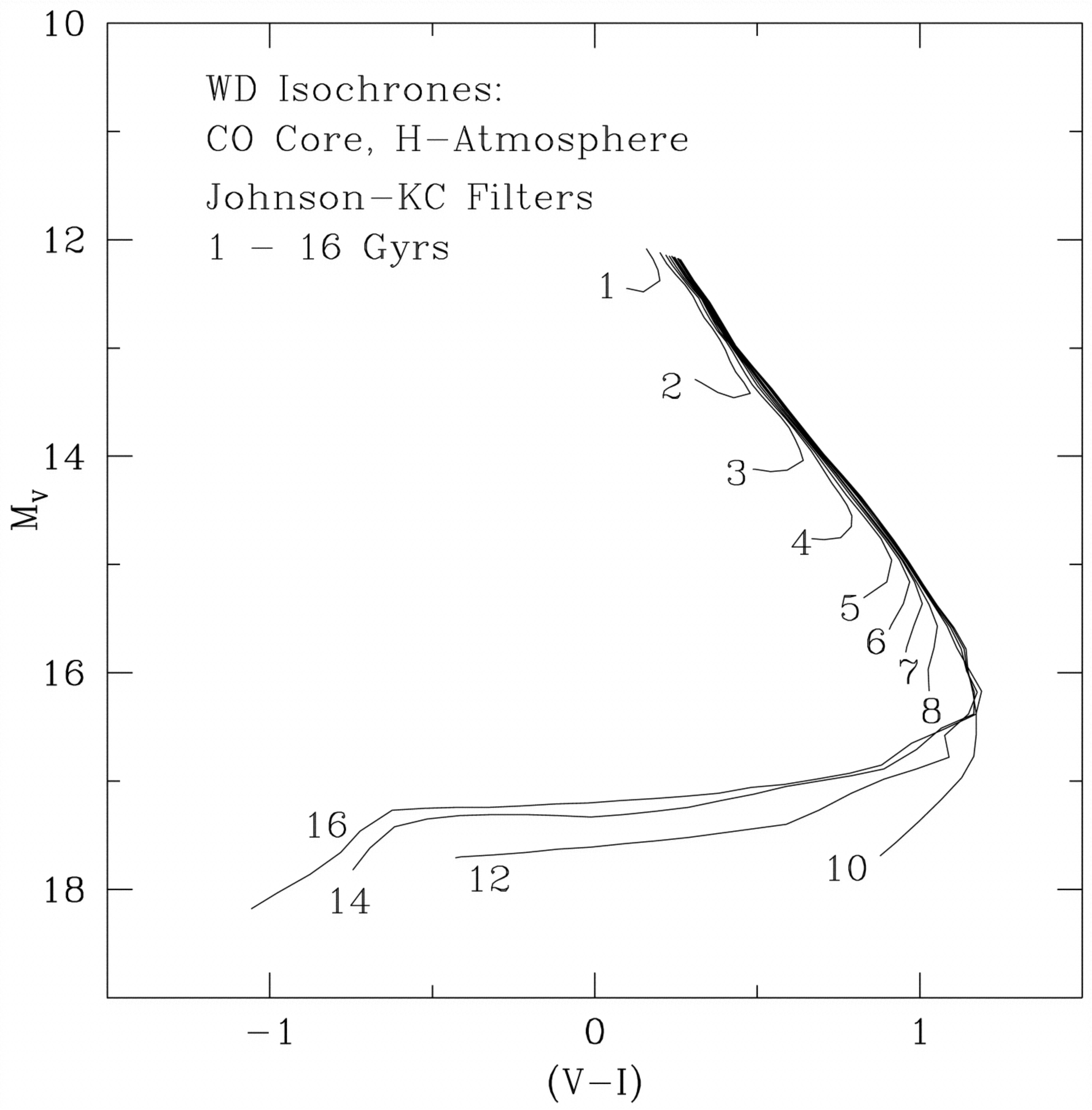}
 \caption{\small Isochrones in Johnson-Kron/Cousins filters describing the turn off along the white dwarf's cooling sequence. Isochrone ages are indicated.
 The image is reproduced from \cite{ri00} by permission of the AAS.}\label{Fig:wd}
\end{figure}

If this limit is exceeded, the structure would collapse to become a completely degenerate system supported 
by the pressure of degenerate neutrons, or a neutron star\index{neutron star}.\\
The white dwarfs have structures which are almost isothermal.
The internal temperature is around $10^{7}$ K (only the outermost layer, which constitutes 1\% of the total thickness, 
is located at a temperature of  $10^{4}$--$10^{5}$ K), and these objects evolve by cooling down at constant radius. 
To describe such cooling sequences in the HR diagram, we must firstly consider that, from the mass-radius relation and the
Stefan-Boltzmann equation, we have:

\begin{equation}
 \log\frac{L}{L_\odot} = 4\log\frac{T}{T_\odot}-\frac{2}{3}\log\frac{M}{M_\odot} + c ,
\end{equation}
meaning that at a given temperature, the more massive a white dwarf is, the less luminous it is. Furthermore, from 
the equation of radiative transport combined with the equation of hydrostatic equilibrium, we get:

\begin{equation}
 L \propto \frac{M_{WD}}{{\rm M}_\odot} \frac{\mu T^{7/2}}{Z(1+X)} ,
\end{equation}
where the brightness of the white dwarf is due to the residual energy of the components of ions  of the structure. 
Finally, comparing this relation to the time derivative of the thermal energy of the ions, a time-temperature 
relation derives, which replaced in the previous one gives:

\begin{equation}
 L_{WD}=L_{0}\left(1+\frac{5t}{2\tau}\right)^{-7/5} ,
\end{equation}
where $\tau$ is the characteristic cooling time for a structure of this type. In light of these laws and of the fact that 
the minimum mass of a white dwarf, according to what already described, is about 0.5 M$_\odot$, a population of 
white dwarfs in a observed colour-magnitude diagram describes a sequence characterised by a ``turn off'' 
that could be used as an indicator of age, but that unfortunately is also very weak and difficult to observe (Fig.~\ref{Fig:wd}).

The final stages of the evolution of stars in the mass range 8 M$_\odot<M<11 $ M$_\odot$ is
quite uncertain. This uncertainty is due to the fact that in this mass range the nucleus of CO formed after the AGB is 
not completely degenerate. In these conditions, therefore, the trigger of the carbon burning happens in an environment 
partially cooled by neutrinos, just the same as during the He-flash. If this occurs, the subsequent evolution is quite 
similar to that of the AGB. Hence, the combustion in a shell of carbon starts, followed by several flashes 
until the complete removal of the degeneration. The similarity is such that we define this phase as a phase of \textit{super-AGB}\index{super AGB}.\\
However, a second scenario considers the possibility that reactions of electron capture become active within the nucleus 
with the effect to remove the degenerate electrons supporting the stellar structure. With the mass of the core being similar 
to the Chandrasekhar limit\index{Chandrasekhar mass}, the structure will collapse to form a neutron star, in a process analogous to that of the core-collapse
supernovae but with much less energy. 

It remains, finally, to describe the final stages for mass stars $M>11$ M$_\odot$. 
In these stars, the high mass ensures the ignition of all the nuclear reactions subsequent to those of hydrogen in a non-degenerate environment, up to the reactions that produce  $^{56}Fe$. At this point, with the formation of a core of iron completely degenerate, the internal temperature has reached values of about $10^{10}$ K, sufficient to activate the process of photo-disintegration of iron nuclei, namely:

\begin{equation}
{\rm  ^{56}Fe }+ \gamma \longrightarrow 13 {\rm ^{4}He }+ 4n
\end{equation}
and very quickly
\begin{equation}
{\rm  ^{4}He} + \gamma \longrightarrow 2p^{+} + 2n .
\end{equation}
These free protons trigger what in astroparticle physics is called the URCA process
\begin{equation}
 p^{+} + e^{-} \longrightarrow n + \nu
\end{equation}
and the support provided by the degenerate electrons is removed. The internal supply of the process leads to the 
collapse of the core. When the density becomes about $10^{14}$ gcm$^{-3}$, the collapse stops and a shock wave 
expels the remaining outer layers, causing a  core-collapse supernova (SN)\index{supernova}. It is important to note that the high flow of 
free neutrons allows the formation, by neutron capture, of r-process elements\index{r-process element}, which become a fundamental tracer for this type of supernovae (which in the old spectral classification are defined as Type II SNe), and that about 99\% of the total energy 
is lost through the emission neutrino. The remnant of this explosion depends, once again, on the initial mass of the star: 
in general, for masses up to 25 M$_\odot$, a neutron star remains after the explosion of a supernova. For masses 
above this limit, we have the formation of a black hole\index{black hole}.

\begin{acknowledgement}
The authors have used as primary references to prepare this chapter the seminal works of Castellani \cite{ca85} and Salaris \& Cassisi \cite{sa05}. It is a pleasure to thank these authors for their exhaustive and illuminating works.
\end{acknowledgement}

%\printindex

%\end{document}

%

\backmatter%%%%%%%%%%%%%%%%%%%%%%%%%%%%%%%%%%%%%%%%%%%%%%%%%%%%%%%
%\appendix
%\include{appendix}
%\include{glossary}
\printindex

%%%%%%%%%%%%%%%%%%%%%%%%%%%%%%%%%%%%%%%%%%%%%%%%%%%%%%%%%%%%%%%%%%%%%%

\end{document}